\documentclass[12pt,preprint]{aastex}

\usepackage{amsmath, amsthm, amssymb}
\usepackage{natbib}
\usepackage{amssymb}
\usepackage{graphicx}
\usepackage{multirow}
\usepackage[normalem]{ulem}
\usepackage{color}
\newcommand{\g}{\gamma}
\newcommand{\tbh}{\hat{t}_b}

\begin{document}

\title{An observational imprint of the Collapsar model of long Gamma Ray Bursts}
\author{\large Omer Bromberg$^1$, Ehud Nakar$^2$, Tsvi Piran$^1$, Re'em Sari$^1$\\
\footnotesize $^1$ Racah Institute of Physics, The Hebrew University, 91904 Jerusalem, Israel\\
\footnotesize $^2$ The Raymond and Berverly Sackler School of Physics and Astronomy,\\
\footnotesize Tel Aviv University, 69978 Tel Aviv, Israel}
%\maketitle
%\begin{document}
%\maketitle
%\begin{affiliations}
%\item Racah Institute of Physics, The Hebrew University, 91904 Jerusalem, Israel
%\item  The Raymond and Berverly Sackler School of Physics and Astronomy,
% Tel Aviv University, 69978 Tel Aviv, Israel
%\end{affiliations}
%

\begin{abstract}
The Collapsar model provides a theoretical framework for  the well
known association between long gamma-ray bursts (GRBs) and
collapsing massive stars. A bipolar relativistic jet, launched at
the core of a collapsing star, drills its way through the stellar
envelope and breaks out of the surface before producing the observed
gamma-rays. While a wealth of observations associate GRBs with the
death of massive stars,  as yet there is no direct evidence for the
Collapsar model itself. Here we show that a distinct signature of
the Collapsar model is the appearance  of a plateau in the duration
distribution of the prompt GRB emission at times much shorter than
the typical breakout time of the jet. This plateau is evident in the
data of all three major satellites. These findings provide an
evidence that directly supports the Collapsar model. Additionally,
it suggests the existence of a large population of choked (failed)
GRBs and that the 2 s duration commonly used to separate Collapsars
and non-Collapasars is inconsistent with the duration distributions
of {\it Swift} and Fermi GRBs and only holds for BATSE GRBs.
\end{abstract}

\section{Introduction}

There  is a long line of evidence connecting long GRBs (LGRBs) to
collapsing massive stars \citep[for recent reviewes
see][]{Woosley06, Hjorth11}. Among them are the association of half
a dozens GRBs with spectroscopically confirmed broad-line Ic
supernovae (SNe), and the identification of "red bumps" in the
afterglows of about two dozens more, which shows a photometric
evidence of underlying SNe. On top of that, the identification of
LGRB host galaxies as highly star forming galaxies and the
localization of the LGRBs in the most active star forming regions
within  those galaxies
\citep{Bloom02,LeFloc'h03,Christensen04,Fruchter06}, provides an
indirect evidence for the connection of LGRBs with massive
stars. The model that provides the theoretical framework of the
LGRB-SN association is known as the Collapsar model
\citep{MacWoos99,MacFadyen01}. According to this model, following
the core collapse of a massive star, a bipolar jet is launched at
the center of the star. The jet drills through the stellar envelope
and breaks out of the surface before producing the observed
gamma-rays. However, although this model is supported
indirectly by the LGRB-SN association, to this date we could not
identify a clear direct observational imprint of the jet-envelope
interaction, thus there is no direct confirmation yet of the
Collapsar model.

In this letter we analyze the expected duration distribution of the
prompt GRB emission from Collapsars. We show that under very general
conditions the time that the jet spends drilling through the star
leads to a plateau in the duration distribution at times much
shorter than the breakout time of the jet. We examine the duration
distribution of all major GRB satellites and find an extended
plateau in all of them over a duration range that is expected for
reasonable jet-star parameters. We interpret these plateaus as the
first identified imprint of the Collapsar model. Under this
interpretation these findings (i) supports the hypothesis of compact
stellar progenitors, (ii) imply the existence of  a large population
of chocked jets that fail to break out of the progenitor star and
(iii) enable us to determine the transition duration that
statistically separates Collapsars from non-Collapsars bursts, and
to show that this time is individual for each satellite. The
separation between Collapsars from non-Collapsars is quantified and
discussed in length in \cite{B11c}. The letter is built accordingly:
In section 2 we briefly review the propagation of a jet inside a
star, and the duration of the prompt GRB emission. In section 3 we
analyze the expected duration distribution of Collapsars. We compare
the expected distribution with the observed one in section 4, and
discuss our findings in section 5.

\section{The propagation of a jet in the stellar envelope}

As the jet propagates in the stellar envelope, it pushes the matter
surrounding it, creating a ``head" of shocked matter at its front.
For typical luminosities the head remains sub relativistic across
most of the star \citep{Matzner03,Zhang03,Morsony07,Mizuta09,B11a},
even though the jet is ejected at relativistic velocities. The jet
breaks out of the stellar surface after \citep{B11b}:
\begin{equation}\label{eq:tB_GRB}
t_b \simeq15~{\rm sec} \cdot \left( \frac{ L_{iso}} {10^{51} {\rm~
erg/sec}}\right)^{-1/3} \left (\frac{\theta}{10^\circ}\right)^{2/3}
\left (  \frac{R_{*}} {5 R_\odot}\right)^{2/3} \left
(\frac{M_{*}}{15M_\odot}\right)^{1/3},
\end{equation}
where $ L_{iso}$ is the isotropic equivalent jet luminosity,
$\theta$ is the jet half opening angle and we have used typical
values for a long GRB. $ R_{*}$ and $M_{*}$ are the radius and the
mass of the progenitor star, where we normalize their value
according to the typical radius and mass inferred from observations
of the few supernovae (SNe) associated with long GRBs. 
While the jet
propagates inside the star the head dissipates most of its energy
and the engine must continuously power the jet in order to support
its propagation. For a successful jet breakout, the engine working
time, $t_e$, must be larger than the breakout time, $t_b$. If $t_e <
t_b$ the jet fails to escape and a regular long GRB is not observed.

Once the jet breaks out from the star it produces the observed
emission at large distances from the stellar surface. Because of
relativistic effects \citep{SariPiran97} the observed duration of
the prompt $\gamma$-rays emission, $t_\gamma$, reflects the time
that the engine operates after the jet breaks out\footnote{
Clearly the engine cannot be working in the same mode  for a time
that is longer than $t_\gamma+t_b$ while if $t_\gamma \gg t_e-t_b$
it is expected to leave a clear signature on the temporal evolution
of the prompt emission, which is not seen \citep{Lazar+09} }:
\begin{equation}
\label{tgamma} t_\gamma = t_{e} -  t_b .
\end{equation}
The distribution of $t_\gamma$ is therefore a convolution of  the
distributions of the engine activity time and the breakout time
combined with cosmological redshift effects.

\section{The duration distribution of Collapsars}\label{sec theory}

Under very general conditions, Eq. \ref{tgamma} results in a flat
distribution of $t_\gamma$ for durations significantly shorter than
the typical breakout time. To see it, consider first a single value
of $t_b$ and ignore, for simplicity, cosmological redshift and
detector threshold effects. The probability that a GRB has a
duration $t_\gamma$ equals, in this case,  to the probability that
the engine work time is $t_\gamma+t_b$. Namely,
\begin{equation}\label{eq:p_g}
p_\gamma(t_\gamma) dt_\gamma=p_{e}(t_b+t_\gamma) dt_\g,
\end{equation}
where $p_\gamma$ is the probability
distribution of observed durations and $p_{e}$ is the probability
distribution of engine working times. Expanding
$p_e(t_b+t_\g)$ around $t_e=t_b$ gives:
\begin{equation}
p_e(t_b+t_\g) = p_e(t_b) + {\cal O}\left(\frac{t_\g}{t_b}\right)
\end{equation}
Thus $p_e(t_b+t_\g) \approx p_e(t_b) =$ const for any $t_\g$ that is
sufficiently smaller than $t_b$. Moreover, if $p_e(t_e)$ is a smooth
function that does not vary rapidly in the vicinity of $t_e \approx
t_b$ over a duration of the scale of $t_b$, then the constant
distribution is extended up to times $t_\g\lesssim t_b$. In the case
of interest  $t_b$ and $t_e$ are determined by different regions of
the star: the breakout time is set by the density and radius of the
stellar envelope at radii $>10^{10}$ cm, while  $t_e$ is determined
by the stellar core properties at radii $<10^8$ cm. The core and the
envelope are weakly coupled \citep{Crowther07} and the engine is
unaware whether the jet has broken out or not. Therefore, it is
reasonable to expect $p_e(t_e)$ to be smooth in the vicinity of $t_e
\approx t_b$ and $p_\gamma (t_\gamma) \approx$ const for $t_\gamma
\lesssim t_b$. In the opposite limit, where $t_b \ll t_\g$, then
$t_\g \approx t_e$, and eq. \ref{eq:p_g} reads:
\begin{equation}\label{tg}
p_\gamma (t_\gamma)  \approx \left\{
\begin{array}{cc}
  p_e(t_b) &  t_\gamma \lesssim t_b   \\
%  F(t_\gamma) dt_\gamma&  t_\gamma \sim t_b   \\
  p_e(t_\gamma) &   t_\gamma \gg t_b
\end{array}
\right. ,
\end{equation}

In reality we expect $t_b$ to vary from one burst to another due to
a variety of progenitor sizes and masses and a scatter in the jet
properties. Moreover, effects such as the dependence of the break
time on the jet luminosity may introduce correlations between $t_e$
and $t_b$. Therefore we consider next a distribution of breakout
times, $p_b(t_b)$, which can generally be correlated with $p_e(t_e)$
(we still ignore cosmological redshift and detector threshold
effects). Since the stellar and jet properties are bounded (e.g.,
there is a minimal progenitor size and maximal jet luminosity), it
implies that within the entire population of bursts there is a
minimal breakout time, $t_{b,min}$. Then eq. \ref{eq:p_g} is
modified to:
\begin{equation}\label{eq P_g2}
    p_\g(t_\g) dt_\g = dt_\g~\int_{t_{b,min}} p_b(t_b) p_e(t_b+t_\g|t_b)dt_b~ ,
\end{equation}
where $p_e(t|t_b)$ is the density distribution of $t_e$ given that
the breakout time is $t_b$. In this case all the arguments presented
above for a single breakout time hold for each value of $t_b$
independently. Therefore, each distribution $p_e(t|t_b)$ has a
plateau at times $t\lesssim t_b$ with a normalization
$p_b(t_b)p_e(t_b|t_b)dt_b$. It can readily be seen that for $t_\g
\lesssim t_{b,min}$ a plateau exists in all the distributions of
engine work times, implying that the right hand side of  Eq. \ref{eq
P_g2} can be written as an integration over plateaus, with different
normalization:
\begin{equation}\label{eq Pgmin}
    p_\g(t_\g \lesssim t_{b,min}) \approx  \int_{t_{b,min}} p_b(t_b) p_e(t_b|t_b)dt_b = {\rm ~const} .
\end{equation}
Thus, $p_\g$ has a plateau at times $t_\g \lesssim t_{b,min}$. For
any reasonable distribution of $p_b$ there is a typical value
$\tbh$, which dominates the contribution to the  integral. Namely,
bursts with $t_\g < t_{b,min}$ are dominated by events with $t_b
\sim \tbh$. When $p_\g$ is a monotonically decreasing function, as
seen in the case of GRBs, then bursts with $t_b \sim \tbh$ dominate
$p_\g$ for any $t_\g \lesssim \tbh$ and the plateau extends up to
$\tbh$.

Finally we take into considerations also the effects of cosmological
redshift and detector threshold. All the quantities that we consider
from here on are observed quantities. Namely, for  bursts at a
redshift z, all durations are stretched by a factor of (1+z),
furthermore all distributions are the observed ones and are affected
by the detector thresholds. For abbreviation we use the same
notations as above. Let the observed redshift distribution be
$p_z(z)dz$, which may be correlated with any other observed
quantity. The observed burst duration distribution is then:
\begin{equation}\label{eq Pgz}
    p_\g(t_\g) = \iint p_z(z) p_b(t_b|z) p_e(t_b+t_\g|t_b,z) dz~ dt_b~~~.
\end{equation}
where $p_e(t|t_b,z)$ is the density distribution of $t_e$ given a
breakout time $t_b$ and redshift z. Following the same
considerations as above, if $p_e(t_b|t_b,z)$ doesn't vary much in
the vicinity of $t_b$ (for every $t_b$ and $z$), then $p_\g$ has a
plateau. Similar reasoning to the one that follows Eq. \ref{eq
Pgmin}, implies that if $p_\g$ is monotonically decreasing function,
then a flat distribution is observed up to $\tbh$, which is the
breakout time whose contribution to the integral in \ref{eq Pgz}
dominates its value at $t_\g<\tbh$. The integral is also expected to
be dominated by bursts from a typical redshift $\hat z$, implying
that the typical intrinsic break time is $\tbh/(1+\hat{z})$.

At long observed durations, $t_\g> \tbh$,  the distribution $p_\g$
can in general depend on all three observed probability
distributions, $p_b$, $p_e$ and $p_z$ and their coupling. However,
if bursts with the same breakout time, $t_b \approx \tbh$, and at
the same redshift, $z \approx \hat z$, dominate the observed
distribution both at $t_\g<\tbh$ and at $t_ \g>\tbh$, then
$p_\g(t_\g \gg \tbh) \propto p_e(t_\g|\tbh,{\hat z})$. Therefore, an
extrapolation of $p_\g(t_\g \gg \tbh)$ to durations shorter than
$\tbh$ is similar to an extrapolation of $p_e(t_e \gg
\tbh|\tbh,{\hat z})$ to a duration $t_e<\tbh$. Namely, it is an
estimate of the number of choked bursts. Now, if bursts with $\tbh$
and $\hat z$ do not dominate the observed distribution at $t_\g >
\tbh$ then, for any reasonable conditions, the monotonically
decreasing  $p_\g(t_\g > \tbh)$ is decreasing less rapidly with
$t_\g$ than $p_e(t_e \gg \tbh|\tbh,{\hat z})$. In that case
extrapolation of $p_\g$ to short durations ($<\tbh$) again provides
a reasonable estimate for the minimal number of choked GRBs.

To conclude, under very general conditions the jet-envelope
interaction in the Collapsar model is predicted to produce a plateau
in the duration distribution of GRBs at short observed durations.
This is true for any breakout time, engine working time and redshift
distributions (including cases where the various distributions are
correlated) as long as the engine working time distribution is
smooth enough. The bursts in the constant section of $p_\g$ are
dominated by a population with an observed breakout time $\tbh$, and
the plateau is extended up to $t_\g \approx \tbh$. Based on the
typical observed GRB parameters (see eq. \ref{eq:tB_GRB}) and given
that a typical GRB is observed at redshift $\approx 2$ we expect
$\tbh \approx 50$ s. Finally, it is well established that at much
shorter durations ($\lesssim 1$ s) the duration distribution
contains a significant population of bursts that are  not associated
with the death of massive stars, which are known as short hard GRBs
(SGRBs)
\citep{Kouveliotou93,Narayan+01,Matzner03,Fox05,Berger05,Nakar07}.
These bursts are non-Collapsars, and the above arguments don't apply
to them. Therefore, when considering the overall  burst duration
distribution we expect a flat section for durations significantly
lower than 50 sec down to the duration where these non-Collapsars
dominate.

\begin{figure}
\includegraphics[width=7in]{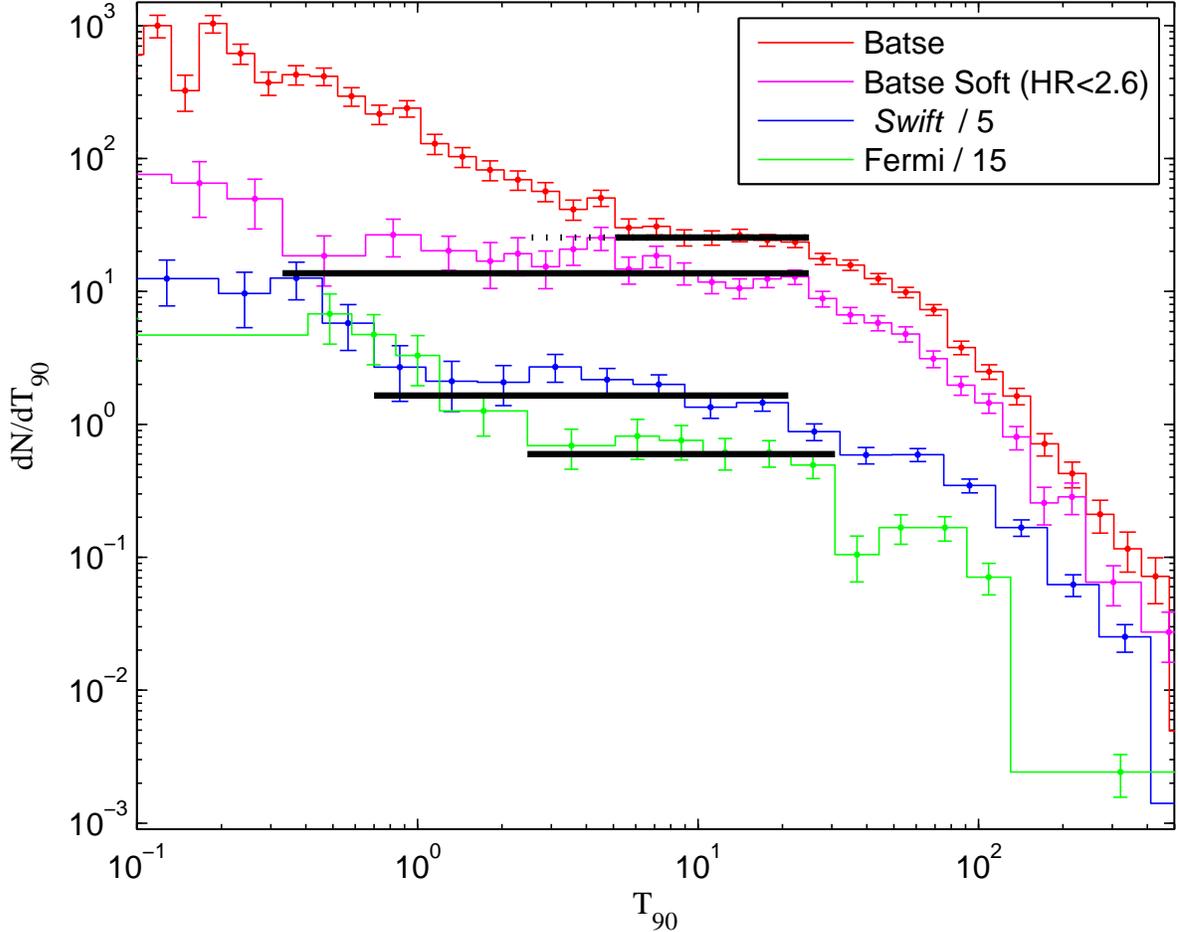}
  \caption{The $T_{90}$ distribution, $dN/dT_{90}$, of BATSE (red), {\it Swift} (blue) and Fermi GBM (green) GRBs.
    Also plotted is the distribution of the soft (hardness ratio $< 2.6$) BATSE bursts (magenta).
    For clarity the {\it Swift} values are divided by a factor of 5 and the Fermi GBM by 15.
    The dotted line that ranges down to $\approx 2$ sec mark the duration range where
    Collapsars constitute more than 50\% of the total number of BATSE GRBs.
    At shorter times the sample is dominated by non-Collapsars.
    Note that the quantity $dN/dT$  is depicted and
    not ${dN}/{d\log T}$ as traditionally shown in such plots \citep[e.g.,][]{Kouveliotou93}.
    The black lines show the best fitted flat interval in each data set:
    $5-25$ sec (BATSE), $0.7-21$ sec ({\it Swift}),  and $2.5-31$ sec (Fermi). The upper limits of this range
    indicate a typical breakout time of a few
    dozens seconds, in agreement with the prediction of the Collapsar model.
    The distribution at times $\gtrsim 100$ sec can be fitted as a power law with an  index
    $-4<\alpha<-3$.
        Soft BATSE bursts  show a considerably longer plateau
    ($0.4-25$ sec), indicating that most of the soft short bursts are in fact
    Collapsars.}
\label{fig.T90_T_tot}
\end{figure}

\section{The observed distribution the prompt GRB durations}

The  observed duration of a GRB is characterized  using
$T_{90}\approx t_\gamma$,  during which  $90\%$ of the fluence is
accumulated.
We use  the data from the three major GRB detectors: BATSE, {\it
Swift} and Fermi GBM. For BATSE we use the current catalog
(04/21/91-05/26/00; containing 2041 bursts). The data of {\it Swift}
is taken from its online archive\footnote
{http://swift.gsfc.nasa.gov/docs/swift/archive/grb\_table}
(12/17/04-08/27/11; containing 582 bursts). Fermi data is extracted
from GCNs using the GRBox website\footnote
{http://lyra.berkeley.edu/grbox/grbox.php} (08/12/08-07/21/11;
containing 194 bursts). Each data set is binned into equally spaced
logarithmic bins, where the minimal number of events per bin is
limited to five \citep{NumRes}. A bin with less than five
events is merged with its neighbor. We use a $\chi^2$ minimization
to look for the longest logarithmic time interval that is consistent
with a flat line within 1 $\sigma$, where the only free parameter is
the normalization. We verify that varying the bin size doesn't
change the length of the plateau by much.

Fig. 1 depicts  the observed distribution of $T_{90}$,
$p_\gamma(T_{90})$, for the three major GRB satellites. Note that we
show here the quantity $p_\gamma(T_{90})=dN/dT$ and not ${dN}/{d\log
T}$ traditionally shown in such plots
\citep[e.g.,][]{Kouveliotou93}. The best fitted flat regions are
highlighted in a solid bold line on top of each distribution. In all
satellites these plateaus range about an order of magnitude in
durations (BATSE 5-25 sec, 3.6/6 $\chi^2$/dof; {\it Swift} 0.7-21
sec, 8.9/7 $\chi^2$/dof; Fermi 1.2-31 sec, 4.1/6 $\chi^2$/dof).
%These results indicate a $\tbh$ value of a few dozens of seconds.
 The extent of the plateau
varies slightly from one detector to another. This is expected given
the different detection threshold sensitivities in different energy
windows (see below).
At the high end of the plateau the $T_{90}$ distribution
decreases rapidly and can be fitted at long durations ($>$100 s) by
a power law with an index, $\alpha$, in the range $-4<\alpha<-3$.

The existence of the plateaus and their duration range ($\sim 2-25$
s) agrees well with the expectation of the Collapsar model. However,
one cannot exclude the possibility that the origin of the observed
flat sections is unrelated to the effect of the jet breakout time
that we discuss above. For example, these plateaus may somehow arise
coincidentally from the combination of two distributions: one
increasing (LGRBs) and one decreasing (SGRBs). In order to test the
hypothesis that the observed plateaus indeed reflect the
distribution of LGRBs, which are Collapsars, we use the fact that
SGRBs are harder \citep{Kouveliotou93}. Restricting the analysis
only to soft bursts should preferentially remove SGRBs from the
sample. Thereby, LGRBs should dominate the duration distribution of
a sample of soft bursts down to durations that are shorter than in
the case of the entire sample. Now, if LGRBs are collapsars and
there duration distribution is flat at short times, then $T_{90}$
distribution of a sample of soft bursts should exhibit a plateau
that extends to shorter durations than the $T_{90}$ distribution of
the whole sample. We present in fig. 1 also a distribution of BATSE
soft bursts (magenta), which are defined as bursts with hardness
ratio\footnote{HR is the fluence ratio between BATSE channels 3
(100-300 keV) and channel 2 (50-100 keV).}, $HR<2.6$, the median
value of bursts with $T_{90}>5$ sec. Remarkably, the plateau in this
sample extends from 25 sec down to $0.4$ sec (15.2/12 $\chi^2$/dof),
over almost two orders of magnitude in duration, compared to the
original 5 sec in the complete BATSE sample. This lands a strong
support to the conclusion  that the observed flat distribution is
indeed indicating on the Collapsar origin of the population. It also
implies that $HR$ is a good indicator that effectively filters out a
large number of non-Collapsars from the GRB sample.

\section{Discussion}

The observed plateaus in all three durtion distributions, and most
notably in the distribution of the soft Batse bursts, provide a
direct support for the Collapsars model for LGRBs. An inspection of
different regions of the observed temporal distribution (Fig. 1),
under the interpretation of the plateau as an imprint of the time it
takes the jet to break out of the envelope, provides further
important information.
\begin{enumerate}

\item{
The end of the plateau and the decrease in the number of GRBs at
long durations, allows us to estimate the typical time it takes a jet to
breakout of the progenitor's envelope. All three distributions are
flat below $\sim 10$ sec in the GRB frame, implying that $\hat t_b
\sim$ a few dozen seconds. This value fits nicely with the canonical
GRB parameters taken in Eq. \ref{eq:tB_GRB}, and provides another
support that the stellar progenitors of Collapsars must be compact
\citep{Matzner03}.}

\item{
The end of the plateau at long durations is dominated by the
distribution of jet breakout times rather than by the engine working
times distribution. Thus, as we discuss in section \ref{sec theory},
an extrapolation of the distribution at long durations, $t_\gamma
\gg \tbh$, to durations shorter than $\tbh$ provides {\it a rough
estimate} of the number of Collapsars with engines that don't work
long enough for their jets to break out. If the jet fails to break
out then it dissipates its energy into the stellar envelope
producing a so called a ``choked" or ``failed" GRB. There have been
numerous suggestions that such a hypothetical population of hidden
``choked GRBs" exists and that these are strong sources of high
energy neutrinos \citep{Eichler99,MW01} and possibly gravitational
waves \citep{Norris03,Daigne07}.

At long durations $p_\gamma(T_{90}\gtrsim 100 {\rm ~ sec})$ can be
fitted well with a power law, $p_\gamma \propto T_{90}^{\alpha}$
with $-4<\alpha<-3$. Extrapolating to $t_e < \hat t_b$ we find that
if $p_e$ continues with this power law to $t_e < \tbh$ there are
many more chocked GRBs than long ones. For example, even if we
extrapolate  this distribution only down to $t_e = \hat t_b/2$ there are still $\sim10$
times chocked GRBs than long GRBs. This prediction is consistent
with the suggestion that shock breakout from these choked GRBs
produces a low luminosity smooth and soft GRB
\citep{Tan01,Wang07,Nakar11}. Indeed the rate of such low luminosity
GRBs is far larger than that of regular long GRBs
\citep{Soderberg06}.}

\item{At short durations, the GRB distribution is dominated by the
non-collapsars SGRBs \citep{Eichler89,Fox05,Berger05,Nakar07}.
They are hard to classify since all their hard energy
properties largely overlap with those of the Collapsars \citep{Nakar07}.
The  least overlap is in the duration distributions and hence, traditionally a
burst is classified as a non-Collapsar if $T_{90} < 2$ sec. Even
though this criterion is based on the duration distribution of BATSE
it is widely used for bursts detected by all satellites.

The fraction of Collapsars at short durations can be estimated by
extrapolating the plateau in the duration distribution. Since there
is also an overlap with the SGRBs at long durations, the real hight
of the plateaus are somewhat lower than what is shown in fig. 1. A
detailed treatment of this issue is presented in \cite{B11c}.
Nevertheless we can use the observed plateaus to obtain a crude
estimation to the duration below which the majority of GRBs are
non-collapsars, by extrapolating the plateaus to the bins in which
Collapsars constitutes 50\% of the bursts. This shows that for BATSE
the transition occurs at $\sim3$ s, while for {\it Swift} it occurs
at $\sim0.7$ s and  at $\sim 1.2$ s for Fermi, although the
statistics in the Fermi data is quite poor. The shift in the
transition time is expected since non-Collapsar bursts are also
harder and different detectors have different energy detection
windows. BATSE has the hardest detection window, making it
relatively more sensitive to non-Collapsar GRBs. {\it Swift} has the
softest detection window making it relatively more sensitive to
Collapsar GRBs.

Thus,  putting the dividing line between Collapsars and
non-Collapsars at 2 sec is statistically reasonable for BATSE bursts. However, our
analysis shows that it is clearly wrong to do so for {\it Swift},
and likely so to Fermi bursts.}

\item{While the difference in the lower limit of the flat ranges is
understood qualitatively in view of the detection windows of the
different detectors, the variance in the upper limit is less
obvious. It may reflect various selection effects in triggering
algorithms. Unfortunately {\it Swift}'s  complicated triggering
algorithm makes is difficult to explore this point quantitatively. A
more interesting possibility is that  it reflects a physical origin,
e.g. that  different satellites probing populations with different
$\hat t_b$.  This could be explored when the statistical sample of
Fermi GBM becomes sufficiently large.
}

\end{enumerate}

To conclude we remark that
it is intriguing that the unique feature discovered here for the
GRB's duration distribution depends just on the simple relation (Eq.
\ref{tgamma}).  This implies that a similar plateau  is expected in
any transient source whose duration is determined  by two unrelated
processes in a similar manner. In particular, in any case in which
one process emits radiation and  another independent one blocks it
for a while.  For example, this would take place in a source
engulfed by dust which will be obscured until the dust is destroyed
by the radiation wave. This could be of importance in interpreting
numerous transient observations, particularly now at the dawn of
astronomy in the time domain.

This research was supported by an ERC advanced research grant, by
the Israeli center for Excellent for High Energy AstroPhysics (TP),
by an ERC and IRG grants, and a Packard, Guggenheim and Radcliffe
fellowships (RS), by the ISF grant No. 174/08 (EN)

%\bibliographystyle{apj}
%\bibliography{myref}{}

\begin{thebibliography}{32}
\expandafter\ifx\csname
natexlab\endcsname\relax\def\natexlab#1{#1}\fi

\bibitem[{{Berger} {et~al.}(2005){Berger}, {Price}, {Cenko}, {Gal-Yam},
  {Soderberg}, {Kasliwal}, {Leonard}, {Cameron}, {Frail}, {Kulkarni}, {Murphy},
  {Krzeminski}, {Piran}, {Lee}, {Roth}, {Moon}, {Fox}, {Harrison}, {Persson},
  {Schmidt}, {Penprase}, {Rich}, {Peterson}, \& {Cowie}}]{Berger05}
{Berger}, E., {et~al.} 2005, \nat, 438, 988

\bibitem[{{Bloom} {et~al.}(2002){Bloom}, {Kulkarni}, \& {Djorgovski}}]{Bloom02}
{Bloom}, J.~S., {Kulkarni}, S.~R., \& {Djorgovski}, S.~G. 2002, \aj,
123, 1111

\bibitem[{{Bromberg} {et~al.}(2011{\natexlab{a}}){Bromberg}, {Nakar}, \&
  {Piran}}]{B11b}
{Bromberg}, O., {Nakar}, E., \& {Piran}, T. 2011{\natexlab{a}},
\apjl, 739, L55

\bibitem[{{Bromberg} {et~al.}(2011{\natexlab{b}}){Bromberg}, {Nakar}, {Piran},
  \& {Sari}}]{B11a}
{Bromberg}, O., {Nakar}, E., {Piran}, T., \& {Sari}, R.
2011{\natexlab{b}},
  \apj, 740, 100

\bibitem[{{Bromberg} {et~al.}(2011{\natexlab{c}}){Bromberg}, {Nakar}, \&
  {Piran}}]{B11c}
{Bromberg}, O., {Nakar}, E., \& {Piran}, T. 2011{\natexlab{c}}, in
preperation

\bibitem[{{Christensen} {et~al.}(2004){Christensen}, {Hjorth}, \&
  {Gorosabel}}]{Christensen04}
{Christensen}, L., {Hjorth}, J., \& {Gorosabel}, J. 2004, \aap, 425,
913

\bibitem[{{Crowther}(2007)}]{Crowther07}
{Crowther}, P.~A. 2007, \araa, 45, 177

\bibitem[{{Daigne} \& {Mochkovitch}(2007)}]{Daigne07}
{Daigne}, F., \& {Mochkovitch}, R. 2007, \aap, 465, 1

\bibitem[{{Eichler} \& {Levinson}(1999)}]{Eichler99}
{Eichler}, D., \& {Levinson}, A. 1999, \apjl, 521, L117

\bibitem[{{Eichler} {et~al.}(1989){Eichler}, {Livio}, {Piran}, \&
  {Schramm}}]{Eichler89}
{Eichler}, D., {Livio}, M., {Piran}, T., \& {Schramm}, D.~N. 1989,
\nat, 340,
  126

\bibitem[{{Fox} {et~al.}(2005){Fox}, {Frail}, {Price}, {Kulkarni}, {Berger},
  {Piran}, {Soderberg}, {Cenko}, {Cameron}, {Gal-Yam}, {Kasliwal}, {Moon},
  {Harrison}, {Nakar}, {Schmidt}, {Penprase}, {Chevalier}, {Kumar}, {Roth},
  {Watson}, {Lee}, {Shectman}, {Phillips}, {Roth}, {McCarthy}, {Rauch},
  {Cowie}, {Peterson}, {Rich}, {Kawai}, {Aoki}, {Kosugi}, {Totani}, {Park},
  {MacFadyen}, \& {Hurley}}]{Fox05}
{Fox}, D.~B., {et~al.} 2005, \nat, 437, 845

\bibitem[{{Fruchter} {et~al.}(2006){Fruchter}, {Levan}, {Strolger},
  {Vreeswijk}, {Thorsett}, {Bersier}, {Burud}, {Castro Cer{\'o}n},
  {Castro-Tirado}, {Conselice}, {Dahlen}, {Ferguson}, {Fynbo}, {Garnavich},
  {Gibbons}, {Gorosabel}, {Gull}, {Hjorth}, {Holland}, {Kouveliotou}, {Levay},
  {Livio}, {Metzger}, {Nugent}, {Petro}, {Pian}, {Rhoads}, {Riess}, {Sahu},
  {Smette}, {Tanvir}, {Wijers}, \& {Woosley}}]{Fruchter06}
{Fruchter}, A.~S., {et~al.} 2006, \nat, 441, 463

\bibitem[{{Hjorth} \& {Bloom}(2011)}]{Hjorth11}
{Hjorth}, J., \& {Bloom}, J.~S. 2011, ArXiv e-prints

\bibitem[{{Kouveliotou} {et~al.}(1993){Kouveliotou}, {Meegan}, {Fishman},
  {Bhat}, {Briggs}, {Koshut}, {Paciesas}, \& {Pendleton}}]{Kouveliotou93}
{Kouveliotou}, C., {Meegan}, C.~A., {Fishman}, G.~J., {Bhat}, N.~P.,
{Briggs},
  M.~S., {Koshut}, T.~M., {Paciesas}, W.~S., \& {Pendleton}, G.~N. 1993, \apjl,
  413, L101

\bibitem[{{Lazar} {et~al.}(2009){Lazar}, {Nakar}, \& {Piran}}]{Lazar+09}
{Lazar}, A., {Nakar}, E., \& {Piran}, T. 2009, \apjl, 695, L10

\bibitem[{{Le Floc'h} {et~al.}(2003){Le Floc'h}, {Duc}, {Mirabel}, {Sanders},
  {Bosch}, {Diaz}, {Donzelli}, {Rodrigues}, {Courvoisier}, {Greiner},
  {Mereghetti}, {Melnick}, {Maza}, \& {Minniti}}]{LeFloc'h03}
{Le Floc'h}, E., {et~al.} 2003, \aap, 400, 499

\bibitem[{{MacFadyen} \& {Woosley}(1999)}]{MacWoos99}
{MacFadyen}, A.~I., \& {Woosley}, S.~E. 1999, \apj, 524, 262

\bibitem[{{MacFadyen} {et~al.}(2001){MacFadyen}, {Woosley}, \&
  {Heger}}]{MacFadyen01}
{MacFadyen}, A.~I., {Woosley}, S.~E., \& {Heger}, A. 2001, \apj,
550, 410

\bibitem[{{Matzner}(2003)}]{Matzner03}
{Matzner}, C.~D. 2003, \mnras, 345, 575

\bibitem[{{M{\'e}sz{\'a}ros} \& {Waxman}(2001)}]{MW01}
{M{\'e}sz{\'a}ros}, P., \& {Waxman}, E. 2001, Physical Review
Letters, 87,
  171102

\bibitem[{{Mizuta} \& {Aloy}(2009)}]{Mizuta09}
{Mizuta}, A., \& {Aloy}, M.~A. 2009, \apj, 699, 1261

\bibitem[{{Morsony} {et~al.}(2007){Morsony}, {Lazzati}, \&
  {Begelman}}]{Morsony07}
{Morsony}, B.~J., {Lazzati}, D., \& {Begelman}, M.~C. 2007, \apj,
665, 569

\bibitem[{{Nakar}(2007)}]{Nakar07}
{Nakar}, E. 2007, \physrep, 442, 166

\bibitem[{{Nakar} \& {Sari}(2011)}]{Nakar11}
{Nakar}, E., \& {Sari}, R. 2011, ArXiv e-prints, 1106.2556

\bibitem[{{Narayan} {et~al.}(2001){Narayan}, {Piran}, \& {Kumar}}]{Narayan+01}
{Narayan}, R., {Piran}, T., \& {Kumar}, P. 2001, \apj, 557, 949

\bibitem[{{Norris}(2003)}]{Norris03}
{Norris}, J.~P. 2003, in American Institute of Physics Conference
Series, Vol.
  686, The Astrophysics of Gravitational Wave Sources, ed. {J.~M.~Centrella},
  74--83

\bibitem[{{Press} {et~al.}(1989){Press}, {Flannery}, {Teukolsky}, \&
  {Vetterling}}]{NumRes}
{Press}, W.~H., {Flannery}, B.~P., {Teukolsky}, S.~A., \&
{Vetterling}, W.~T.
  1989, {Numerical recipes in C. The art of scientific computing}, ed. {Press,
  W.~H., Flannery, B.~P., Teukolsky, S.~A., \& Vetterling, W.~T. } (Cambridge:
  University Press)

\bibitem[{{Sari} \& {Piran}(1997)}]{SariPiran97}
{Sari}, R., \& {Piran}, T. 1997, \apj, 485, 270

\bibitem[{{Soderberg} {et~al.}(2006){Soderberg}, {Kulkarni}, {Nakar}, {Berger},
  {Cameron}, {Fox}, {Frail}, {Gal-Yam}, {Sari}, {Cenko}, {Kasliwal},
  {Chevalier}, {Piran}, {Price}, {Schmidt}, {Pooley}, {Moon}, {Penprase},
  {Ofek}, {Rau}, {Gehrels}, {Nousek}, {Burrows}, {Persson}, \&
  {McCarthy}}]{Soderberg06}
{Soderberg}, A.~M., {et~al.} 2006, \nat, 442, 1014

\bibitem[{{Tan} {et~al.}(2001){Tan}, {Matzner}, \& {McKee}}]{Tan01}
{Tan}, J.~C., {Matzner}, C.~D., \& {McKee}, C.~F. 2001, in American
Institute
  of Physics Conference Series, Vol. 586, 20th Texas Symposium on relativistic
  astrophysics, ed. {J.~C.~Wheeler \& H.~Martel}, 638--640

\bibitem[{{Wang} {et~al.}(2007){Wang}, {Li}, {Waxman}, \&
  {M{\'e}sz{\'a}ros}}]{Wang07}
{Wang}, X.-Y., {Li}, Z., {Waxman}, E., \& {M{\'e}sz{\'a}ros}, P.
2007, \apj,
  664, 1026

\bibitem[{{Woosley} \& {Heger}(2006)}]{Woosley06}
{Woosley}, S.~E., \& {Heger}, A. 2006, \apj, 637, 914

\bibitem[{{Zhang} {et~al.}(2003){Zhang}, {Woosley}, \& {MacFadyen}}]{Zhang03}
{Zhang}, W., {Woosley}, S.~E., \& {MacFadyen}, A.~I. 2003, \apj,
586, 356

\end{thebibliography}

\end{document}